Excitation functions of the $^{nat}$Ta(p,x)$^{178m2}$Hf and $^{nat}$W(p,x)$^{178m2}$Hf reactions at energies up to 2600 MeV

Yu. E. Titarenko[1], V. F. Batyaev[1], K.V. Pavlov[1], A.Yu. Titarenko[1], V. M. Zhivun[1,2], M. V. Chauzova[1], A.V. Ignatyuk[1,3], S.G. Mashnik[4], S. Leray[5], A. Boudard[5], J.-C. David[5], D. Mancusi[5], J. Cugnon[6], Y. Yariv[7], K. Nishihara[8], N. Matsuda[8], H. Kumawat[9], A.Yu. Stankovsky[10]

[1] NRC Kurchatov Institute "Institute for Theoretical and Experimental Physics", Moscow, Russia
[2] NRNU MEPhI (Moscow Engineering Physics Institute), Moscow, Russia
[3] Institute of Physics and Power Engineering, Obninsk, Russia
[4] Los Alamos National Laboratory, Los Alamos, USA
[5] CEA, Saclay, France
[6] University of Liege, Belgium
[7] Soreq NRC, Yavne, Israel
[8] JAEA, Tokai, Japan
[9] BARC, Mumbai, India
[10] SCK•CEN, Belgium

Abstract

Due to potential level of energy intensity $^{178m2}$Hf is an extremely interesting isomer. One possible way to produce this isomer is irradiation of $^{nat}$Ta or $^{nat}$W samples with high energy protons. Irradiation of $^{nat}$Ta and $^{nat}$W samples performed for other purposes provides an opportunity to study the corresponding reactions. This paper presents the $^{178m2}$Hf independent production cross sections for both targets measured by the gamma-ray spectrometry method. The reaction excitation functions have been obtained for the proton energies from 40 up to 2600 MeV. The experimental results were compared with calculations by various versions of the intranuclear cascade model in the well-known codes: ISABEL, Bertini, INCL4.5+ABLA07, PHITS, CASCADE07 and CEM03.02. The isomer ratio for the $^{nat}$Ta(p,x) $^{178m2}$Hf reaction is evaluated on the basis of the available data.

Irradiations of the thin sample-targets of $^{nat}$Ta and $^{nat}$W with protons in the energy range 40-2600 MeV have been performed at the ITEP accelerator during the period from September 1, 2006 to August 31, 2009, within the framework of the ISTC Project #3266. Irradiation conditions and the measured independent and cumulative yields of the main reaction products are presented in Refs. [1 - 3]. About 600 gamma- and alpha-spectra have been measured for both $^{nat}$Ta and $^{nat}$W, from the analysis of which 882 cross sections were estimated for $^{nat}$Ta and 1013 cross sections were obtained for $^{nat}$W. These data were presented in the form of 173 residual excitation functions for $^{nat}$Ta and 193 excitation functions for $^{nat}$W. However, on account of an essential background in the region of gamma-ray lines corresponding to $^{178m2}$Hf, the yield of this isomer has not been determined.

To estimate the yields of $^{178m2}$Hf the measurements of gamma-spectra for the early irradiated samples of $^{nat}$Ta and $^{nat}$W were continued during 2012 – 2014. Preliminary results were published in Ref. [4]. The increased time after irradiation provides a significant decrease in the radioactivity background because of the natural decay of the reaction products with short half-lives, and that allows us to identify the $^{178m2}$Hf by its distinctive gamma lines quite confidently. Two gamma-ray spectrometers with HP Ge detectors of the GC2518 type and the DSA1000 analyzers were used for the spectra measurements. The absolute efficiency of each spectrometer was calibrated by means of the validated gamma-sources: $^{207}$Bi, $^{153}$Gd, $^{22}$Na, $^{241}$Am, $^{228}$Th, $^{152}$Eu, $^{139}$Ce, $^{137}$Cs, $^{133}$Ba, $^{113}$Sn, $^{109}$Cd, $^{88}$Y, $^{60}$Co, $^{57}$Co, $^{54}$Mn, which were certified by the VNIIM State metro-logical center. An example of such calibration is shown in Fig. 1. The background spectrum of the laboratory room, in which the measurements were carried, is shown in Fig. 2.

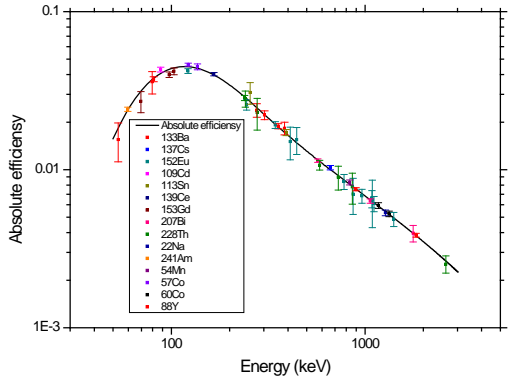
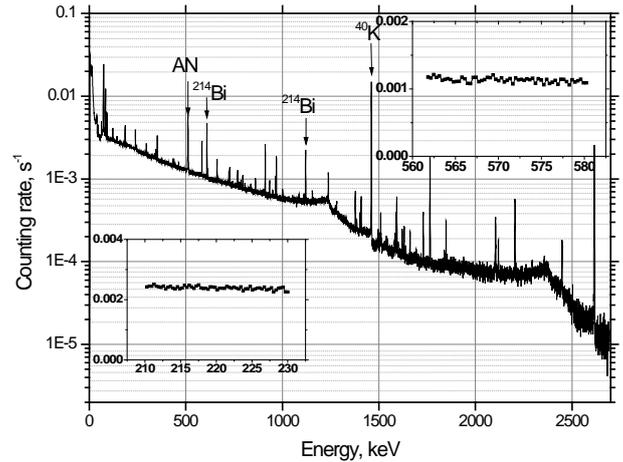

Fig. 1. The efficiency of ND gamma-ray spectrometer.

Fig. 2. The laboratory background spectrum. The measuring time was 14 days.

The $^{178m2}$Hf yield was identified in accordance with the intensity of its gamma-lines: 213.4 keV (81.4%), 216.7 keV (64.6%) and 574.2 keV (88.0 %) [5]. Other gamma-lines were not considered because of their lower yields (<20%) or the overlapping of their energies with the gamma-peaks from other reaction products. As a rule we did not analyze the gamma-lines with energies less than 100 keV.

The measured gamma-spectrum of the $^{nat}$Ta and $^{nat}$W samples for about 5 years after irradiation by 600 MeV and 800 MeV protons is shown in Figs. 3-4. Enlarged plots of the gamma-peaks used for the $^{178m2}$Hf identification are shown in the corresponding insets of this Figs.

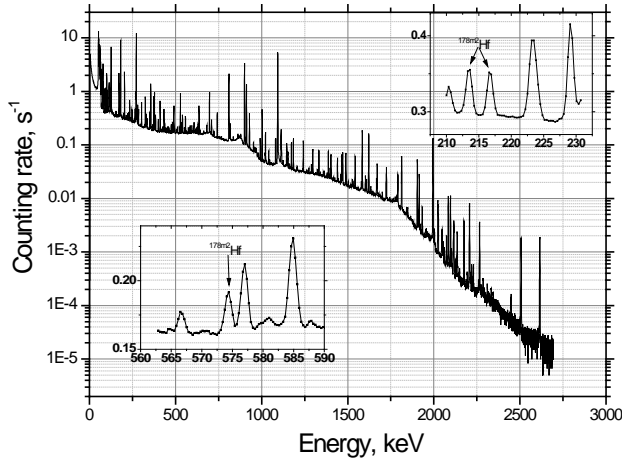
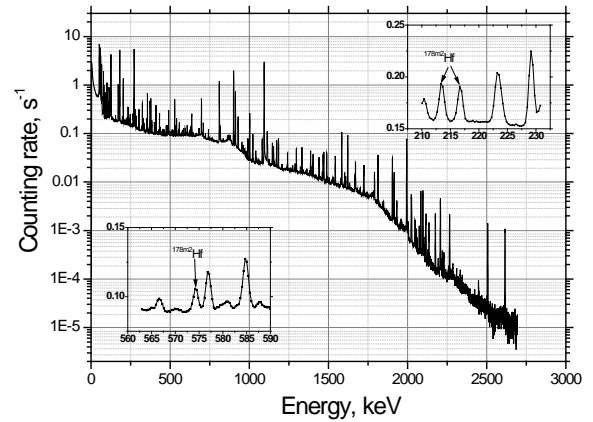

Fig. 3. The measured gamma spectrum of the $^{nat}$Ta sample for about 5 years after irradiation with 600 MeV protons. The measuring time was ~ 7 days.

Fig. 4. The measured gamma spectrum of the $^{nat}$W sample for about 5 years after irradiation with 800 MeV protons. The measuring time was ~ 10 days.

The methodology for determining the cross sections of radioactive reaction products was by means of the gamma-ray spectrometry described in details in Refs. [1-3]. Isobaric chain fragment with mass number A = 178 is shown in Fig. 5 [5].

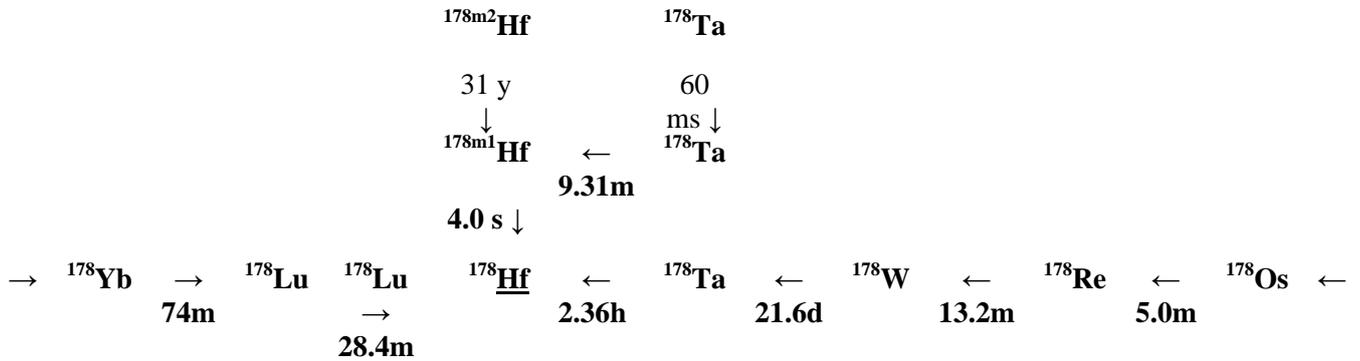

Fig. 5. $^{178}$Hf chain.

Because the $^{178m2}$Hf isomer with the time of life $T_{1/2}$ = 31 y is formed in nuclear reactions only without any contribution from the decay of neighboring residuals, the formulas for calculated independent production rates of *i*-th γ-line can be presented as

$$R_i^{ind} = \frac{A_i}{N_{Tag} \cdot \eta_i \cdot \varepsilon_i \cdot \lambda} \cdot \frac{1}{t_{irr}} \qquad (1),$$

where $R_i^{ind}$ is reaction rate production for *i*-th γ-line $^{178m2}$Hf, $A_i$ is count rate for *i*-th γ-line $^{178m2}$Hf (taking into account "cooling" time and the absorption of γ-rays in the sample which calculated with data from [6]), $N_{Tag}$ is the number of nuclei in the sample, $\eta_i$ is the absolute yield of γ-line, $\varepsilon_i$ is the absolute efficiency of the γ-spectrometer at the analyzed energy, $\lambda$ is the decay constant, $t_{irr}$ is the irradiation time (because $t_{irr} \ll T_{1/2}^{178m2 Hf}$ the correction for decay $^{178m2}$Hf during the irradiation of the sample is not taken into account).

The mean independent reaction rate $^{178m2}$Hf were calculated with a rather simple formula.

$$\overline{R}^{ind} = \frac{\sum_1^3 R_i^{ind} \cdot W_i}{\sum_1^3 W_i}, \quad \text{where} \quad W_i = \frac{1}{\left(\Delta R_i^{ind}\right)^2}. \qquad (2),$$

where $\overline{R}_i^{ind}$ is the mean rate reaction production $^{178m2}$Hf, $\Delta R_i^{ind}$ is uncertainty rate reaction production for *i*-th γ-line $^{178m2}$Hf

The $^{178m2}$Hf cross section production is calculated from the expressions

$$\sigma^{ind} = \frac{\overline{R}^{ind}}{\hat{\Phi}_{st}}, \qquad (3),$$

where $\sigma_i^{ind}$ is independent cross section $^{178m2}$Hf, $\Phi_{st}$ is the proton flux, the data given in Ref. [3].

Uncertainties $\Delta R_i^{ind}$, $\Delta \overline{R}_i^{ind}$ and $\Delta \sigma^{ind}$ were calculated in according to [7]

The cross sections obtained for $^{178m2}$Hf production are given for both targets in Table 1 together with the corresponding uncertainties.

Table 1. Measured cross sections for the $^{nat}$Ta(p,x)$^{178m2}$Hf and $^{nat}$W(p,x)$^{178m2}$Hf reactions

| Proton energy (MeV) | Cross section and its uncertainty (mb) | |
|---|---|---|
| | $^{nat}$Ta | $^{nat}$W |
| 44 | 0.041 ± 0.012 | − |
| 68 | 0.075 ± 0.022 | − |
| 98 | 0.102 ± 0.029 | − |
| 149 | 0.208 ± 0.028 | 0.102±0.028 |
| 248 | 0.189±0.022 | 0.092±0.021 |
| 399 | 0.273 ± 0.026 | 0.191 ± 0.044 |
| 599 | 0.316±0.022 | 0.204±0.015 |
| 799 | 0.290±0.020 | 0.213±0.013 |
| 1199 | 0.278±0.018 | 0.170±0.019 |
| 1598 | 0.215±0.029 | 0.134±0.044 |
| 2605 | 0.257±0.029 | 0.218±0.034 |

The measured cross sections are shown in Figs. 6 and 7, in comparison with the calculated independent and cumulative cross sections for the $^{172}$Hf production. Calculations are performed for various versions of the intranuclear cascade model realized in the well-known codes: ISABEL, Bertini, INCL4.5+ABLA07, PHITS, CASCADE07 and CEM03.02. A brief discussion of these codes was given in Refs. [1 − 3], and a more detailed consideration of their approximations can be found in the original publications refereed in [3]. All codes calculate the residual nuclide production, but, unfortunately, they do not split the nuclide cross sections over the isomeric states.

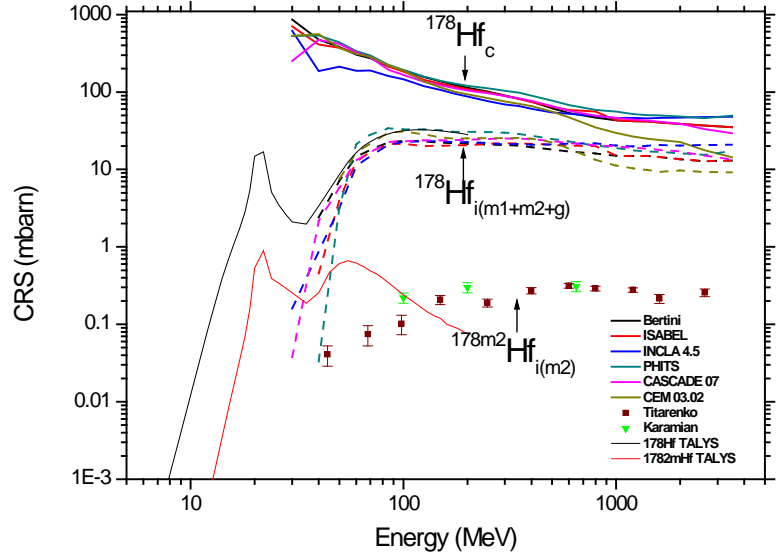

Fig. 6. Calculated cross sections for $^{178}$Hf $_c$, $^{178}$Hf $_{i(m2+m1+g)}$ and experimental data of the excitation function for $^{nat}$Ta(p,x)$^{178m2}$Hf $_{i(m2)}$ reaction.

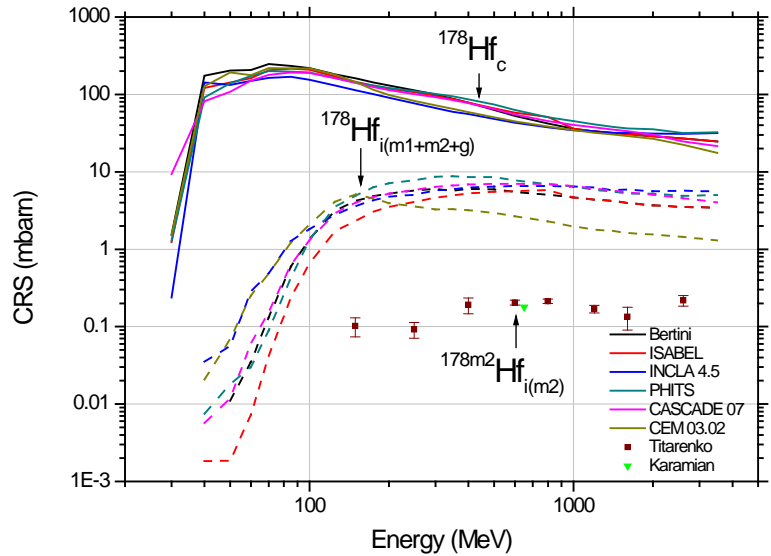

Fig. 7. Calculated cross sections for $^{178}$Hf $_c$, $^{178}$Hf $_{i(m2+m1+g)}$ and experimental data of the excitation function for $^{nat}$W(p,x)$^{178m2}$Hf $_{i(m2)}$ reaction.

In addition, Figs. 6 and 7 present the experimental data which was obtained earlier at JINR, Dubna by Karamian et al. [8]. Despite the fact that in Ref. [8] a thick target (a = 2 cm) had been irradiated, the results agree within the limits of experimental errors for the proton energy of 660 MeV, since the energy loss was only ~ 7%. Discrepancy for energies of 200 MeV and 100 MeV is because of a significant loss of energy (up to stop of the beam).

Fig. 6 includes also the calculations of $^{nat}Ta(p,x)^{178}Hf$ and $^{nat}Ta(p,x)^{178m2}Hf$ cross sections in the energy region below 200 MeV performed with the TALYS code [9]. This code attempts to estimate the split of nuclide-production cross sections over the isomeric states. However, the schemes of low-lying levels should be known for such calculations. Unfortunately, for $^{178}Hf$ the low-lying levels are estimated not well below 2.1 MeV only, while the spins of many levels around the m2-isometric state (2.446 MeV) are not identified yet [5]. In such conditions there are no reasons to hope on a high accuracy of the calculated isomeric cross-sections. The calculated yields for $^{178}Hf$ agree reasonably with the transport-code results at energies above 50 MeV, and a large spread of results arises at the lower energies only, that is connected with the approaches simulating the α-particle production in the preequilibrium processes [9]. The calculated cross sections for the $^{178m2}Hf$ production disagree rather strongly with the experimental data for the whole energy region.

For many applications, an analysis of the isomeric ratios can be more useful than the cross-section consideration. To obtain such ratios we used the TALYS nuclide cross sections in the energy range below 200 MeV and the calculated cross sections averaged over all transport-code data at higher energies. The obtained isomeric rations are shown in Fig. 8 together with the TALYS results. To take into account the uncertainties of cross section estimations, the uncertainties for the isomer ratios were assumed twice as large as the relative uncertainties of experimental data.

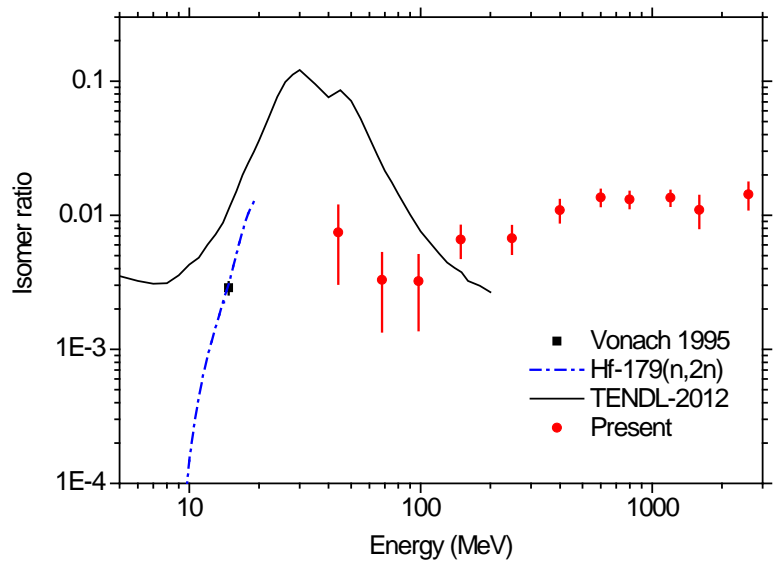

Fig. 8. Estimated isomer ratios for the $^{181}Ta(p,x)^{178m2}Hf$ reaction compared with the TENDL-2012 and the low-energy data for the $^{179}Hf(n,2n)^{178m2}Hf$ reaction.

In neutron-induced reactions, the yield of isomer $^{178m2}Hf$ ($J^{\pi}=16^+$) was analyzed in Refs. [10-12]. For the neutron energy of 14.8 MeV, the estimated yield of $^{178m2}Hf$ in the reaction $^{179}Hf(n,2n)$ corresponds to the value of 6.33 ± 0.28 mb [11]. Since the total cross section for the $^{179}Hf(n,2n)$ reaction at this energy is equal to 2200 ± 150 mb, the value of the isomer ratio is equal ~ 0.003 with an uncertainty about 10%. On the other hand, the empirical systematics of isomeric ratios for neutron-induced reactions provides a value of 0.001 with an uncertainty of more than 50% for the $16^+$ spin case [11].

Theoretical calculations [12] of the isomeric ratio for the high-spin states in the neutron energy up to 20 MeV indicate a systematic growth of the isomeric ratio with an increase of energies. Such growth looks confirmed also by the TALYS results for the energies below 30 MeV, but the absolute values of the isomer ratios are overestimated about ten times in these calculations. On the basis of the experimental data, we can conclude that a more probable behavior of the isomer ratio corresponds to a smooth transition from the low-energy values of the (n,2n) reaction to the decreasing part at the energies from 25-30 to about 80 MeV and the following growth of the isomer ratio to the high-energy value about 0.014±0.002.

The evaluated value of the isomer ratio can be used for a more serious simulation of the level scheme around the m2-isomer. Such scheme will be certainly very interesting for the calculations of $^{178m2}$Hf production in other reactions including the reactions on the $^{nat}$W sample. Accumulation of new experimental data on the high-spin isomer production is certainly important for a further development of nuclear reaction models, too.

## ACKNOWLEDGEMENTS
The authors would like to thank the ITEP accelerator team for the long-time participation in irradiation of various samples. The authors appreciated also very much the support received from the ISTC and IAEA CRP projects, as well as from the current pilot project of the National Research Center "Kurchatov Institute". Part of the work performed at LANL was carried out under the auspices of the National Nuclear Security Administration of the U.S. Department of Energy. Last but not least, we thank Dr. Roger L. Martz for a careful reading of our manuscript and useful suggestions on its improvement.